\documentstyle[epsf]{l-aa}
\def\v{\vskip 2mm}
\def\vv{\v\v}

\def\kms{km s${}^{-1}$ }
\def\kmsm{km s${}^{-1}$ Mpc${}^{-1}$ } 
\def\tfr{{Tully-Fisher relation }}

\makeatletter

\@ifundefined{epsfbox}{\@input{epsf.sty}}{\relax}                       %AASTeX
\def\eps@scaling{.95}                                                   %AASTeX
\def\epsscale#1{\gdef\eps@scaling{#1}}                                  %AASTeX
\def\plotone#1{\centering \leavevmode                                   %AASTeX
\epsfxsize=\eps@scaling\columnwidth \epsfbox{#1}}                       %AASTeX
\begin{document}

   \thesaurus{03         % A&A Section 3: Ectragalactic Astronomy
              (12.04.3;  % Cosmology: distance scale,
               11.04.1;  % Galaxies: distances and redshifts,
               11.09.2;  % Galaxies: interactions,
               11.07.1;  % Galaxies: general,
               11.09.4;  % Galaxies: ISM
               13.19.1)} % Radio lines: galaxies
   \title{Effects of Galaxy Interaction on the Tully-Fisher Relation:}

   \subtitle{CO vs HI Linewidths}

   \author{Y. Tutui and Y. Sofue } 
%          \inst{1}

   \offprints{Y. Tutui }

   \institute{ Institute of Astronomy, University of Tokyo, 
               Mitaka, Tokyo 181, Japan; E-mail: tutui@mtk.ioa.s.u-tokyo.ac.jp
             }

   \date{received July 1, 1996 /  accepted June 9, 1997}

   \maketitle

   \begin{abstract}

We investigate effects of galaxy-galaxy interaction on the
Tully-Fisher relation. 
The HI linewidth in interacting galaxies is significantly
broader than the CO linewidth,
and the HI to CO linewidth ratio  
is proportional to the strength of interaction.
This provides different distances to galaxies measured 
by the Tully-Fisher relation with the CO and HI linewidths.
Distances derived from the HI linewidths
are 62\% larger than those derived from the CO linewidths
for  strongly interacting galaxies,
and  25\% larger 
for  weakly interacting galaxies.
We argue that the CO-line Tully-Fisher relation will be more reliable 
to measure the distances of interacting galaxies as well as galaxies
in rich clusters.

   \end{abstract}

%
%________________________________________________________________

\section{Introduction}

The HI Tully-Fisher relation
has been the most successful and widely applied tool to measure
distances to galaxies
(e.g., Tully \&  Fisher 1977; Aaronson et al. 1986; Pierce \& Tully 1988).
CO linewidth has been also used in the 
\tfr instead of HI linewidth.
HI linewidth almost coincides with CO linewidth for 
galaxies in the Coma cluster and other nearby clusters 
(Dickey \& Kazes 1992), for field galaxies (Sofue 1992; 
Sch\"oniger \& Sofue 1994) and for Virgo cluster galaxies (Sch\"oniger \&
Sofue 1997).

Because the beamsize of CO observations is much sharper than that
for HI,
we are able to resolve individual galaxies at higher redshift,
and  avoid contamination by other galaxies in one beam
for CO observations.
Observations for the CO-line Tully-Fisher relation have been performed 
using the Nobeyama Radio Observatory (NRO) 45-m telescope
(HPBW = 15''),
% at ${}^{12}CO (J = 1-0)$ line.
and CO linewidths have been obtained for galaxies at redshift 
$cz \sim$ 29,000 \kms (Sofue et al. 1996),
at which the CO beamsize is still small enough to distinguish 
galaxies.

HI gas extends far from the center of galaxies beyond the optical radius, 
while it is deficient in the central region (Bosma 1981). 
On the other hand, the molecular gas 
is known to be more concentrated 
in a better correlation with the optical disk
(Young \& Scoville, 1982). 
Since the atomic and molecular gases in galactic disk are distributed 
separately in radius (Sofue et al.1995; Honma et al.1995),
the HI linewidths may be more strongly disturbed by galaxy-galaxy 
interaction, which is inevitable in rich clusters of galaxies 
and at high redshifts.
Interacting galaxies are generally excluded from a sample of 
nearby galaxies for the Tully-Fisher relation.
However, when we measure distances to farther galaxies,
we may overlook features of interaction and overestimate the distances.

In this paper we examine the correlation between CO and HI linewidths
of nearby galaxies
and discuss the  Tully-Fisher relation for interacting
galaxies.

%
%__________________________________________________________________

\section{Data and Sample Selection}

In this study the data of CO linewidths are taken from 
Young et al. (1995),
who observed the ${}^{12}$CO ($J = 1- 0$) emission line of 
300 field galaxies using the 14-m telescope of 
the Five College Radio Astronomy 
Observatory (FCRAO) (HPBW = 45''). 
Among them the CO emission was detected at multiple positions 
for 103 galaxies.
We excluded the following galaxies from our analysis;
(1) galaxies for which we could not obtain linewidths 
due to weak or strange position-velocity diagrams,
(2) face-on galaxies ($ i < 30 {}^{\circ}$) 
in order to minimize the effect of inner velocity dispersion.
As the result of these selection we selected 60 galaxies.
We classified these galaxies into 17 interacting galaxies
and 43 isolated galaxies.
Then we classified the interacting galaxies into three subclasses
using the interaction class (hereafter IAC) defined by Dahari (1985).
Although the Dahari's IAC is classified into 6 groups based on features of 
interaction,
we classified them into three classes 
due to the small sample.
Our classification based on Dahari's IAC is as follows;
\begin{itemize}
   \item[(1)] weakly interacting galaxies 
              (hereafter WIG, Dahari's IAC = 2,3),
   \item[(2)] strongly interacting galaxies 
              (hereafter SIG, Dahari's IAC = 4,5),
   \item[(3)] mergers  (Dahari's IAC = 6).
\end{itemize}
According to Dahari's classification, IAC = 1 stands for 
isolated galaxies.
The classification for individual galaxies based on Dahari's method
is described in Table 3a and 3b, as well as Column 9 in Table 1.
The galaxies which Dahari classified are marked with asterisks.

The CO linewidths are obtained from position-velocity diagrams in 
the literature (Young et al. 1995).
%The uncertainty of the values are less than 20 km/s.
The HI linewidth and inclination of galaxies are taken from 
Huchtmeier \& Richter (1989).
Among HI data in the catalog, we selected the data
observed at Arecibo 1000-ft, Effelsberg 100-m,
NRAO Green Bank 300-ft, 
Jodrell Bank 250-ft and Parks 64-m telescopes
in order to keep the quality of data set.  

Linewidths are defined as the full width at 20\% of
maximum intensity of the global profile.
In order to correct for an inclination effect, 
we adjust to an edge-on orientation by: 
${W}_{\rm i}  =   W / \sin ~ i,$ 
where $W$ is the observed linewidth, $i$ is the inclination and 
${W}_{\rm i}$ is the linewidth corrected for the inclination. 
Total magnitude, the Galactic and internal extinction corrections and 
radial velocity corrected
to the Galactic Standard of Rest are taken from the Third Reference
Catalog of Bright Galaxies(RC3) (de Vaucouleurs et al. 1991).   

%%%%%%%%%%%%%%%%%%%%%%%%%%%%%%%
%   Table 1
%%%%%%%%%%%%%%%%%%%%%%%%%%%%%%%
\vv

\begin{table*}

{\bf Table 1.}  Linewidths and Distances for the Interacting Galaxies\\ 

\begin{tabular}{l c c r c r r c l} \hline
Galaxy & ${W}_{\rm i~CO}$ & ${W}_{\rm i~HI}$ & $ \Delta {W}_{\rm i}$ &
 ${W}_{\rm HI}/{W}_{\rm CO}$ & ${D}_{\rm CO}$ & ${D}_{\rm HI}$ &
${D}_{\rm HI}/{D}_{\rm CO}$ & IAC \\ 
 & (km/s) & (km/s) & (km/s) &        & (Mpc)& (Mpc) &    & \\ \hline
NGC  520 & 475 & 410 & $ -65$ &  0.863 & 41.9 & 33.6 & 0.86 &${6}^{*}$\\
NGC  660 & 419 & 360 & $ -59$ &  0.859 & 24.0 & 19.2 & 0.86 &6\\
NGC  772 & 546 & 602 &   56 &  1.103 & 30.4 & 35.2 & 1.10 &3\\ 
NGC 1961 & 731 & 910 &  179 &  1.245 & 63.2 & 87.6 & 1.24 &4\\
NGC 2146 & 531 & 595 &   64 &  1.121 & 23.1 & 27.3 & 1.12 &6\\
NGC 2798 & 266 & 403 &  137 &  1.515 & 19.7 & 36.6 & 1.51 &${5}^{*}$\\
NGC 3034 & 302 & 292 & $ -10$ &  0.967 &  5.5 &  5.2 & 0.97 &4\\
NGC 3169 & 835 & 692 & $-143$ &  0.829 & 54.1 & 40.8 & 0.82 &3\\
NGC 3627 & 436 & 451 &   15 &  1.034 &  8.8 &  9.3 & 1.03 &${3}^{*}$\\
NGC 3628 & 383 & 483 &  100 &  1.261 &  7.9 & 11.2 & 1.26 &3\\
NGC 4038 & 459 & 602 &  143 &  1.312 & 18.6 & 27.9 & 1.31 &5\\ 
NGC 4088 & 348 & 407 &   59 &  1.170 & 11.9 & 15.1 & 1.17 &3\\
NGC 4631 & 307 & 355 &   48 &  1.156 &  4.1 &  5.1 & 1.16 &3\\
NGC 5054 & 317 & 418 &  101 &  1.319 & 13.7 & 20.8 & 1.32 &4\\
NGC 5194 & 340 & 626 &  286 &  1.841 &  4.9 & 12.2 & 1.84 &${4}^{*}$\\
NGC 5713 & 274 & 370 &   96 &  1.350 & 13.6 & 21.4 & 1.35 &3\\ 
NGC 6217 & 358 & 485 &  127 &  1.355 & 22.3 & 35.2 & 1.35 &2\\
\hline 
\end{tabular}

\vspace{4mm}
\noindent
{\bf Column 1}: Galaxy name. 
{\bf Column 2, 3}: CO- and HI-linewidth 
corrected for the inclination, respectively. 
{\bf Column 4}: Linewidth difference between HI and CO 
defined by ${W}_{\rm i~HI} - {W}_{\rm i~CO}$.
{\bf Column 5}: HI-to-CO linewidth ratio. 
{\bf Column 6, 7}: Distances derived from 
the Tully-Fisher relation with the CO and HI linewidths, respectively.
{\bf Column 8}: HI-to-CO distance ratio.
{\bf Column 9}: Interaction class (IAC). 
The IAC which Dahari classified is marked with an asterisk.

\end{table*}

%%%%%%%%%%%%%%%%%%%%%%%%%%%%
%   3. RESULT
%%%%%%%%%%%%%%%%%%%%%%%%%%%%

\section{Results}

Figure 1 shows a plot of the CO linewidths (${W}_{\rm i~ CO}$)  versus
the HI linewidths (${W}_{\rm i~HI}$) for 17 interacting galaxies and
43 isolated galaxies.
The CO and HI linewidths are corrected for the inclination.
The CO and HI linewidths approximately coincide with each other 
for most of isolated galaxies (open circles). 
On the other hand, in most cases of interacting galaxies (filled symbols), 
the HI linewidths are 
clearly broader than those of CO.

Among the interacting galaxies, the differences between the CO and HI line
widths in the SIG (filled squares) are larger than in the WIG 
(filled circles). 
Table 1 shows the number of galaxies, mean value and 
standard deviation of the ratio for each class. 
Here we excluded three galaxies which have small linewidth less
than 200 \kms, namely NGC 598, NGC 3893 and NGC 2976, 
from the statistics, because errors for small linewidth 
would amplify the linewidth ratio.
The sample of mergers is small so that we cannot discuss their 
statistics.

%%%%%%%%%%%%%%%%%%%%%%%%%%%%
% Fig. 1
%%%%%%%%%%%%%%%%%%%%%%%%%%%%
\begin{figure}[htb]
\vspace{9cm}
\includegraphics{5462.f1.ps}
      \caption{ 
The CO linewidths versus the HI linewidths 
corrected for inclination. 
Isolated galaxies are marked by open circles.
Interacting galaxies are marked by filled symbols;
mergers by filled triangles, SIG 
by filled squares, and  WIG by filled circles.
The subclasses of the interacting galaxies are explained in Section 2.
              }
   \end{figure}

%%%%%%%%%%%%%%%%%%%%%%%%%%%%
% Fig. 2
%%%%%%%%%%%%%%%%%%%%%%%%%%%%
\begin{figure}[b]
\vspace{5cm}
\includegraphics{5462.f2.2.ps}
\caption{ 
Histograms of $W_{\rm HI}/W_{\rm CO}$  
for isolated galaxies (top), weakly interacting galaxies (middle) and
strongly interacting galaxies (bottom).
              }
\end{figure}

%%%%%%%%%%%%%%%%%%%%%%%%%%%%
%   Table.2
%%%%%%%%%%%%%%%%%%%%%%%%%%%%

\begin{table*}
{\bf Table 2.} HI-to-CO Linewidth Ratio and Distance Ratio\\

\begin{tabular}{l c c c c c c} \hline
  & & & & & & \\          
Classification & IAC${}^{1}$    & Number &          
 $\overline{{W}_{\rm HI}/{W}_{\rm CO}}$ & $\sigma$ & 
 $\overline{{D}_{\rm HI}/{D}_{\rm CO}}$ & $\sigma$ \\   \hline
Isolated Galaxies ${}^{2}$         & 1   & 40 &  1.01 & 0.10 & 1.02 & 0.11\\
Weakly Interacting Galaxies(WIG)   & 2,3 &  8 &  1.16 & 0.16 & 1.25 & 0.26\\
Strongly Interacting Galaxies(SIG) & 4,5 &  6 &  1.37 & 0.27 & 1.62 & 0.47\\ 
Mergers${}^{3}$                    & 6   &  3 &  (0.95) & (0.12) & (0.93) & (0.18)\\
\hline 

\end{tabular}

\noindent
${}^{1}$ The IAC is defined by Dahari (See Appendix). \\
${}^{2}$ Three galaxies with  CO linewidths less than  200
\kms are excluded.\\
${}^{3}$ The number of mergers is small so that we do not discuss the 
statistics of  mergers.\\

\end{table*}

%%%%%%%%%%%%%%%%%%%%%%%%%%%%%%

Figure 2 shows histograms of ${W}_{\rm HI}/{W}_{\rm CO}$ 
for isolated galaxies, weakly and strongly interacting galaxies.
The interacting galaxies have obviously larger values of the ratio
than the isolated galaxies.
The ratio ${W}_{\rm HI}/{W}_{\rm CO}$ is proportional to 
strength of the interaction. 
This trend is caused by the tidal force of 
the galaxy - galaxy interaction.
The CO gas is tightly confined to the luminous stellar disk,
while the HI gas extends even beyond the optical disk.
Therefore, the HI gas is likely to be disturbed by the tidal force,
and the HI linewidth is broadened.

In order to compare distances derived from the CO and HI linewidths, 
we measured the distances using the Tully-Fisher relation in B-band.
We assume that the same Tully-Fisher relation is adopted for
the CO and HI linewidths,
because there is no significant difference between the CO and HI
linewidths for the isolated galaxies and the same relationship is
better to compare between them.
The B-band Tully-Fisher relation which we adopted is 
given by Pierce \& Tully (1992),
\begin{equation}
{M}_{\rm B} = - 7.48 ({\rm log}{W}_{\rm i} -2.5) -19.55,
\end{equation}
where ${M}_{\rm B}$ is the B-band absolute magnitude.
In order to examine which of the linewidths is reliable for 
interacting galaxies, we plotted recession velocity versus distances  
derived from the Tully-Fisher relation with the CO and HI linewidths, 
and measured the Hubble constants.
We compared the Hubble constants with those for the isolated galaxies.
Figures 3 show the velocity-distance diagrams,
namely the Hubble diagrams for interacting galaxies 
and isolated galaxies.
The solid line and the dotted line are regression lines of 
the CO and HI data, respectively.
The Hubble constants  derived from the CO and HI linewidths  
for the isolated
galaxies are ${H}_{0} = 60.8 \pm 6.9 $ and 60.9 $\pm$ 5.7 \kmsm, 
respectively.
The CO and HI data give a consistent value of ${H}_{0}$ for the 
isolated galaxies.
Here, the errors are due to the dispersion within the sample.
On the other hand,
the Hubble constants derived from the CO and HI linewidths 
for the interacting galaxies are 65.2 $\pm$ 11.5 and 
54.0 $\pm$ 9.3 \kmsm, respectively.
This indicates  that the HI linewidths  
are broadened with the significance level of 89 \%, 
and cause larger differences in ${H}_{0}$ estimates 
for the interacting galaxies.

%%%%%%%%%%%%%%%%%%%%%%%%%%%%
% Fig. 3
%%%%%%%%%%%%%%%%%%%%%%%%%%%%
\begin{figure}[tb]
\vspace{7cm}
\includegraphics{5462.f3.ps}
\caption{ 
Hubble diagrams for interacting galaxies (top)
and isolated galaxies (bottom),
radial velocities referred to the Galactic Standard of Rest (from RC3)
versus distances derived from the Tully-Fisher relation.
Symbols are the same as in Fig. 1.
Distances derived from the CO linewidths are plotted by filled symbols 
and those derived from the HI linewidths are  plotted
by open symbols.
The solid and dotted lines indicate the regression lines of the CO and 
HI data, respectively.
              }
\end{figure}

%%%%%%%%%%%%%%%%%%%%%%%%%%%%
%    4.Discussion
%%%%%%%%%%%%%%%%%%%%%%%%%%%
\section{Discussion and Summary}
  
We discuss the effect of interaction on the
Tully-Fisher relation.
The relationship between the HI-to-CO linewidth ratio and 
the HI-to-CO distance ratio is given as,

\begin{eqnarray}
\frac{{D}_{\rm HI}}{{D}_{\rm CO}} & = &  {\left(\frac{{W}_{\rm i~HI}}{{W}_{\rm i~CO}}\right)}^{k/5}
\end{eqnarray}
where $k$ is the slope of the Tully-Fisher relation ($k = 7.48$).
Table 2 gives the obtained values of the ratio.
In order to estimate the error, 
we assume that the error in the linewidths
is $\pm$ 15 \kms, and the error in the slope of the 
Tully-Fisher relation is $\pm$ 0.50.
We combine them with the dispersion of the sample.
Then we find that the HI-to-CO distance ratio is 1.25 $\pm$ 0.28 
for the WIG,
and 1.62 $\pm$ 0.48 for the SIG.
Distance derived from the HI linewidths is, thus, 
found to be 62\% larger 
than that derived from the CO linewidths  for SIG.
When we observe distant galaxies, we may overlook tidal features,
because they are faint.
We suggest that the CO-line  Tully-Fisher relation 
will be more reliable to measure
the distances of galaxies for especially distant galaxies.
The same should apply for galaxies 
in rich clusters, 
where galaxy-galaxy interaction is far more frequent and inevitable.

Finally we mention that
the effect of interaction should be taken into account 
not only in the Tully-Fisher relation but also in the dynamics 
and evolution of spiral galaxies.
The present method of the ${W}_{\rm CO}$ versus ${W}_{\rm HI}$ comparison 
may give a clue to reveal dynamical properties of galaxies 
in the era and regions where galaxy-galaxy interaction would have 
a significant effect.

We summarize our results as follows:
\begin{itemize}
 \item[(1)] HI linewidths are larger than CO linewidths for
interacting galaxies, 
and the linewidth ratio ${W}_{\rm HI}/{W}_{\rm CO}$ is 
proportional to the strength of interaction.
 \item[(2)] Distances derived from   
the HI Tully-Fisher relation are  25\% larger for the weakly
interacting galaxies and 62\% larger for the strongly interacting
galaxies than that derived from the CO Tully-Fisher relation. 
 \item[(3)] Therefore, the CO Tully-Fisher relation would give more
reliable distances for interacting galaxies.
This implies that the CO Tully-Fisher relation will give 
better distance measurement for galaxies inside rich
clusters,
where the tidal interaction is inevitable.
\end{itemize}

%%%%%%%%%%%%%%%%%%%%%%%%%%
%   Appendix
%%%%%%%%%%%%%%%%%%%%%%%%%%
\vv
\noindent
{\bf Appendix}

\v
We classified the interacting galaxies
into 3 classes, WIG, SIG and mergers.
The classification is based on Dahari's method (Dahari 1985).
Dahari classified 167 systems of interacting and asymmetric galaxies
into six groups.
%They defined the dimensionless 'interaction class'(IAC) as an integer,
%which grows with the interaction effect on the galaxies, 
%as seen projected on the sky.
%IAC = 1 is assigned to isolated symmetric galaxies, 
%and IAC = 6 is assigned to severely 1 spirals or overlapping 
%systems.
%The IAC is determined by the degree of asymmetry of the galaxy,
%by the distance and size of a companion, and by the presence of
%connecting arms between the pair members.
%The presence of a nearby companion (with a comparable redshift)
%almost certainly imposes a significant tidal force on the galaxy.
%It is not certain whether radial flows are present when a galaxy 
%is still symmetric despite the presence of a close companion.
The IAC classifications of single- and double- galaxy systems 
are described in Table 3a and 3b, respectively.
All galaxies in our sample are listed in the tables.

%%%%%%%%%%%%%%%%%%%%%%%%%%%%
%   Table.3a
%%%%%%%%%%%%%%%%%%%%%%%%%%%%
\begin{table}[h]
{\bf Table 3a.} Definition of Interacting Class (IAC) of Single Galaxies\\

\begin{tabular}{l l c } \hline
IAC            & Description    &  Galaxy    \\        
               &                &  NGC        \\  \hline
1 .....&  Symmetric (isolated)     &           \\
2 .....&  Slightly asymmetric, diffuse extensions & 6217   \\
3 .....&  Asymmetric, extended arms  & 4088   \\
4 .....&  Distorted, out of shape  & 1961    \\
5 .....&  Strongly disordered &  $-$        \\
6 .....&  Aftermath          &  660,2146    \\
\hline 
\end{tabular}
%\vspace{4mm}

\end{table}

%%%%%%%%%%%%%%%%%%%%%%%%%%%%
%   Table.3b
%%%%%%%%%%%%%%%%%%%%%%%%%%%%

\begin{table}[h]
{\bf Table 3b.} Definition of Interaction Class(IAC) for Pair Galaxies\\

%\begin{footnotesize}
\begin{tabular}{l c l  c l  c l } 
\hline
    &   \multicolumn {6}{c}{Companion Size}\\
\cline {2-7}
             & \multicolumn {2}{c}{Same} & \multicolumn {2}{c}{$\sim
1/2$} &  \multicolumn {2}{c}{Small}\\ 
\cline{2-3}
\cline{4-5}
\cline{6-7}       
Separation   &    {\scriptsize IAC} & {\scriptsize NGC} & {\scriptsize  IAC} & 
{\scriptsize NGC} &  {\scriptsize IAC} & {\scriptsize NGC  } \\  
\hline
{\scriptsize Large, no contact} & 3 & 3169         &  2 &  $-$   &  1 &  $-$     \\
                  &   & 3627${}^{*}$ &    &        &    &           \\
                  &   & 3628         &    &        &    &           \\
                  &   & 5713         &    &        &    &           \\
{\scriptsize Large, connected}  & 4 & 3034         &  3 &  $-$   &  2 &  $-$     \\
{\scriptsize Small, no contact} & 4 &  $-$         &  4 &  $-$   &  3 & 772      \\
                  &   &              &    &        &    & 4631     \\
{\scriptsize Small, connected}  & 5 & 2798${}^{*}$ &  4 & 5194${}^{*}$   &  4 & 5054  \\
                  &   & 4038         &    &        &    &           \\
{\scriptsize Overlap}           & 6 & 520${}^{*}$       &  5 &  $-$   & 4 &  $-$     \\   
\hline 
\end{tabular}
%\end{footnotesize}
\end{table}

%%%%%%%%%%%%%%%%%%%%%%%%%%%
%  References
%%%%%%%%%%%%%%%%%%%%%%%%%%
\newpage
\noindent
{\bf References}

\begin{description}

\item Aaronson M., Bothun G., Mould J., et al.,1986, ApJ 302, 536

\item Bosma A., 1981, AJ 86, 1825

\item Dahari O., 1985, ApJS 57, 643

\item de Vaucouleurs G., de Vaucouleurs A., Corwin H.G. Jr., et al., 1991, 
Third Reference Catalogue of Bright Galaxies.
Springer-Verlag, New York (RC3) 

\item Dickey J.M., Kazes I., 1992, ApJ 393, 530

\item Honma M., Sofue Y., Arimoto N., 1995, A\&A 304, 1

\item Huchtmeier W.K., Richter O.-G., 1989, A General Catalog of HI
 Observation of Galaxies. Springer-Verlag, New York

\item Pierce M.J., Tully R.B., 1988, ApJ 330, 579

\item Sch\"oniger F., Sofue Y., 1994, A\&A 283, 21 

\item Sch\"oniger F., Sofue Y., 1997, (in press) 

\item Sofue Y., 1992, PASJ 44, L231

\item Sofue Y., Honma M., Arimoto N., 1995, A\&A 296, 33

\item Sofue Y., Sch\"oniger F., Honma M., Tutui Y.,et al., 1996, PASJ
48, 657

\item Tully R.B., Fisher J.R., 1977, A\&A 54, 661 

\item Young J.S., Scoville N., 1982, ApJ 258, 467

\item Young J.S., Xie S., Tacconi L., et al., 1995, ApJS 98, 219

\end{description}

\end{document}